# Environmental Performance, Financial Constraint and Tax Avoidance Practices: Insights from FTSE All-Share Companies

**Probowo Erawan Sastroredjo [1,2], Marcel Ausloos [1,3,4,5,*], and Polina Khrennikova [1,6]**

[1] School of Business, University of Leicester, Leicester LE2 1RQ, UK; pes13@leicester.ac.uk (P.E.S.); ma683@leicester.ac.uk (M.A.); p.khrennikova@utwente.nl (P.K.)

[2] Department of Management, Parahyangan Catholic University, Bandung 40164, Indonesia

[3] Group of Researchers Applying Physics in Economy and Sociology (GRAPES), B-4031 Liège, Belgium

[4] Faculty of Economics and Business Administration and Department of Theoretical and Computational Physics Babeș-Bolyai University, 400084 Cluj-Napoca, Romania

[5] Department of Statistics and Econometrics, Bucharest University of Economic Studies, 010552 Bucharest, Romania

[6] Financial Engineering Group, Faculty of Behavioural, Management, and Social Sciences (BMS), University of Twente, 7522 NB Enschede, The Netherlands

* Correspondence: marcel.ausloos@uliege.be

**Abstract**

Through its initiative known as the Climate Change Act (2008), the Government of the United Kingdom encourages corporations to enhance their environmental performance with the significant aim of reducing targeted greenhouse gas emissions by the year 2050. Previous research has predominantly assessed this encouragement favourably, suggesting that improved environmental performance bolsters governmental efforts to protect the environment and fosters commendable corporate governance practices among companies. Studies indicate that organisations exhibiting strong corporate social responsibility (CSR), environmental, social, and governance (ESG) criteria, or high levels of environmental performance often engage in lower occurrences of tax avoidance. However, our findings suggest that an increase in environmental performance may paradoxically lead to a rise in tax avoidance activities. Using a sample of 567 firms listed on the FTSE All Share from 2014 to 2022, our study finds that firms associated with higher environmental performance are more likely to avoid taxation. The study further documents that the effect is more pronounced for firms facing financial constraints. Entropy balancing, propensity score matching analysis, the instrumental variable method, and the Heckman test are employed in our study to address potential endogeneity concerns. Collectively, the findings of our study suggest that better environmental performance helps explain the variation in firms' tax avoidance practices.

## 1. INTRODUCTION

The UK is encountering a significant issue with corporation tax avoidance, primarily stemming from its tax system's intricacies and multinational companies' ability to shift profits across borders to lessen their tax obligations in the UK [0] (as still found on https://www.parliament.uk/globalassets/documents/lords-information-office/2015/hl-annual-report-2013-14.pdf (accessed on 24 December 2024)). Numerous companies use tax avoidance to reduce their tax liabilities by utilising legally available avenues [1]. Although tax avoidance remains a major concern for the UK government, it persistently rolls out initiatives to grant tax relief to firms that earnestly invest in environmentally friendly projects. For instance, the UK government has established programmes to promote environmental, social, and governance (ESG) principles, particularly those focused on environmental sustainability. These programmes enhance firms' operational efficiency while minimising harmful waste production. Various types and sizes of companies can qualify for designated tax incentives and support schemes (This information can be seen at https://www.gov.uk/green-taxes-and-reliefs (accessed on 27 December 2024)). The primary objective is to encourage companies to undertake more environmental initiatives, stimulate green innovation, and exhibit a greater commitment to sustainability. Nevertheless, research by Souguir et al. [2] suggests that while these measures will motivate companies to adopt greener practices, they may also influence their tax strategies.

Considering growing concerns regarding rising tax avoidance practices in the UK, our study examines the potential link between improved environmental performance and corporate tax avoidance. Numerous studies have investigated the factors influencing corporate tax avoidance, attributing it to various characteristics at the firm level. For instance, elements such as economies of scale, foreign operations, and company size significantly affect the extent of tax avoidance [3,4]. Moreover, ownership structure [5] and compensation agreements [6,7] drive corporate tax avoidance behaviours. Despite this, there is a scarcity of research focusing on how environmental performance influences tax avoidance.

Companies pursuing aggressive environmental objectives often face increased initial expenses. These costs arise from investments in sustainable technologies, implementing eco-friendly processes, and measures to reduce carbon emissions [8–10]. To alleviate some of these financial pressures, certain companies might engage in tax avoidance strategies to lessen their tax liabilities, ultimately seeking to enhance cash flow or profitability [11,12]. Social and environmental initiatives safeguard shareholder interests while boosting financial performance [13,14]. Nevertheless, these sustainable practices might not deliver immediate financial gains, mainly if such investments come at the cost of more profitable opportunities [15,16]. Consequently, managers might lean towards aggressive tax avoidance strategies to reduce tax liabilities, improve short-term profitability, and conceal economic vulnerabilities [11,17]. Thus, our paper posits that strong environmental performance correlates positively with corporate tax avoidance activities.

Our study investigates the relationship between enhanced environmental performance and increased corporate tax avoidance. Our findings support new evidence and practical implications. Indeed, analysing a dataset of 1004 firm-year observations from 2014 to 2022 reveals that an increase in environmental practices correlates with greater tax avoidance. Furthermore, the study indicates that this increase in tax avoidance is more significant among firms facing financial constraints. The research employs entropy balancing, propensity score matching, Heckman selection bias correction, and an instrumental variable approach to address potential endogeneity issues. Our approach also considers several alternative measures to confirm the robustness of the documented findings. The follow-up tests yield consistent results with the baseline findings, reinforcing the conclusion that the outcomes of our study are robust and minimise endogeneity concerns.

Our research enriches the existing literature in multiple ways. Firstly, it contributes to the studies examining the factors driving corporate tax avoidance [18–20]. For instance, [18] indicate that individual top executives significantly influence corporate tax avoidance. Na and Yan [19] link tax avoidance practices to linguistic factors, and [20] demonstrate that higher corporate social responsibility (CSR) performance correlates with reduced tax avoidance levels. Our research contributes to the existing literature bydemonstrating that a strong commitment to environmental initiatives significantly elevates the probability of firms participating in tax avoidance practices, contrary to Lanis and Richardson’s findings [20]. Companies with superior environmental performance may utilise tax avoidance to alleviate the financial burden associated with such activities. Secondly, corporate environmental activities have recently become a primary concern for shareholders, government regulators, employees, and customers [21]. In analysing the economic impact of environmental performance, Hassel et al. [22] find a negative correlation between environmental performance and firms’ market value. Therefore, our research expands the discussion by demonstrating that environmental performance significantly enhances tax avoidance practices. Finally, some previous studies [23–25] report a positive relationship between greater financial constraints and tax avoidance.

Our study contributes to the existing body of knowledge by confirming that heightened financial constraints are a significant factor influencing firms' tax avoidance strategies. We enrich the existing literature on the interplay between corporate social responsibility (CSR) and firm-specific variables, particularly focusing on investments in environmental activities that often require additional funding beyond what is necessary for sustaining day-to-day operations. Lastly, this study aims to explore the complex relationship between environmental, social, and governance (ESG) performance and tax behaviour, with a specific focus on the UK market (FTSE All Shares) over the recent period from 2014 to 2022. While our study specifically examines UK companies, its findings have broader implications within a global context, especially given recent initiatives mandating ESG disclosures for publicly listed firms. Our paper hopefully contributes to the international literature on the relationship between CSR (and its dimensions) and corporate tax behaviour. It complements prior studies on other markets, such as [26] on Tunisia, [27] on Indonesian tech companies, Australian and US- focused research in [28,29], studies on Chinese companies [30,31], and analyses by [32] on European markets and by [33] on French companies.

Section 2 provides a review of the relevant literature, outlines the theoretical framework, and develops the research hypotheses. Section 3 details the research design and methodology employed in the study. Section 4 presents the empirical findings and offers a discussion of the results. Finally, Section 5 concludes the study, highlighting key implications and potential avenues for future research.

## 2. LITERATURE REVIEW, THEORY AND HYPOTHESIS DEVELOPMENT

### *2.1. The Cash Flow Theory*

The cash flow theory suggests that tax avoidance helps corporations to increase their cash flow by employing effective tax planning strategies to generate cash flow savings, businesses can leverage these funds as a potential source of financing when confronting financial constraints [34]. This is viable because corporate income tax is a non-discretionary expense imposed on profitable companies. While the exact amount of income tax depends on the company's operating jurisdiction laws, managers can employ different tax strategies to lower corporate tax payments.

Following the theory and its behavioural implications, managers may adjust reported cash flows in response to financial constraints and the associated risks and uncertainties related to sustainability investments undertaken [35]. Reduced tax liabilities enable corporations to maintain liquidity for operational activities, capital investments, and shareholder returns. As such, organisations facing financial constraints may resort to tax avoidance strategies to mitigate economic pressures. Moreover, the benefits derived from tax avoidance can be reinvested into environmental initiatives, thereby establishing a positive feedback loop. Corporations might utilise funds saved as a result from tax avoidance practices as a funding source for sustainability initiatives, aiming to improve their environmental, social, and governance (ESG) scores, particularly as they continue to face governmental pressure to do so. The studies by [36,37] demonstrate that tax planning behaviour among Chinese enterprises is positively associated with the mitigation of financial constraints and the improvement of firms' liquidity. Sun et al. [36] further highlight that tax planning is more pronounced in high-growth sectors where firms are more likely to face capital shortages. Similarly, Kabir et al. [38] empirically confirm, using a global dataset, the conjecture that financial pressures negatively impact companies engaged in environmental investments and eco-innovation. As such, companies often face difficult decisions regarding the short-term and long-term maximisation of returns to their shareholders and balancing the demands of other stakeholders on sustainability targets.

Collectively, these findings inform our research hypotheses: (i) there is a positive association between ESG initiatives and tax planning to minimise cash shortages, and (ii) this relationship is further amplified for companies experiencing or beginning to experience financial pressures.

### *2.2. Environmental Performance and Tax Avoidance*

Carroll's pyramid model remains a seminal framework in understanding corporate social responsibility (CSR) [39], as outlined by Ma et al. [39]. Carroll [40] originally conceptualised CSR as encompassing economic, legal, ethical, and discretionary expectations, hierarchically ranking these responsibilities in the order of economic, legal, ethical, and philanthropic responsibilities, as discussed in [41]. Carroll [42] further refined this framework, highlighting these four dimensions as integral components of CSR. Recently, the measurement of CSR through environmental, social, and governance (ESG) scores has introduced a new dimension, emphasising environmental sustainability and activities as a critical addition to CSR, reflecting its evolving scope in contemporary business practices.

The correlation between cash flow and environmental performance is intricately associated with liquidity and financial planning. Engagement in environmentally sustainable activities or investments may lead to considerable expenditure. Organisations that invest in or implement these initiatives must strategically manage their resources, as a portion of their budget will be designated for tax obligations and adherence to environmental activities. Furthermore, this transition may result in short-term financial pressures, particularly for companies shifting towards more sustainable practices or those with high pollution levels necessitating substantial environmental investments (e.g., see recent global evidence in [38]). Firms experiencing financial pressures often lack the necessary resources to invest in cleaner technologies, resulting in negative environmental effects, such as increased carbon emissions [43].

The relationship between environmental performance and tax avoidance is found to be inconsistent in the academic literature. Some research indicates that involvement in CSR reduces tax avoidance through enhanced transparency and ethical behaviour [28,29,44–47]. Conversely, some scholars argue that CSR activities can facilitate aggressive tax strategies [30,48–50]. Recent insights into ESG factors further complicate this discussion. While some evidence suggests that a focus on ESG could lead to increased tax avoidance for financial flexibility [51], other research posits that strong environmental performance may lower the risk of tax avoidance by fostering ethical accountability [2,27,32].

Companies committed to ambitious environmental performance encounter considerable initial costs from investments in sustainable technologies and eco-friendly processes [8,10,52]. The financial pressures arising from these undertakings may incentivise tax avoidance strategies to enhance cash flow and sustain profitability [11,12]. Moreover, while environmental initiatives are intended to protect shareholder interests and foster long-term financial performance, their lack of immediate financial returns may compel managers to adopt aggressive tax strategies that minimise liabilities and boost short-term earnings and cash flow [15–17].

Consequently, our study hypothesis (H1) is as follows:

**Hypothesis 1.** *High environmental performance is positively associated with corporate tax avoidance avoidance practices.*

*2.3. The Moderating Role of Financial Constraints*

Based on insights from previous research, it is recognised that cash management significant influences a company's financial constraints [53,54]. Companies facing high financial constraints cannot secure external funding for their operations, investments, and growth opportunities. As a result, these companies often resort to aggressive tax planning [23], which leads to tax avoidance. Furthermore, ESG initiatives, particularly those related to environmental investments, require substantial and reliable funding. However, there is a concern that this could prompt companies to avoid tax to meet stakeholders' expectations regarding environmental issues [55].

Based on those findings, our study hypothesis (H2) is as follows:

**Hypothesis 2.** *Financial constraints strengthen the positive association between environmental intensity and tax avoidance.*

## 3. RESEARCH DESIGN AND METHODOLOGY

*3.1. Data and Sample Collection*

This study draws on financial data from Refinitiv Eikon, with 1234 firm-year observations for 567 companies indexed from FTSE All Shares. The sample period ranges from 2014 to 2022, aligning with previous research by Albitar et al. [56]. The selection of this timeline is a direct response to recent policies and updates concerning climate change, particularly the implementation of mandatory disclosure requirements for greenhouse gas (GHG) emissions (Scope 1 and Scope 2) in the UK in 2013. Following the empirical method of [57,58], our study excludes firms in the finance (SIC codes 4000-4999) and utility (SIC codes 6000-6999) sectors. Indeed, financial sectors, including banks and insurance companies, function under rigorous financial reporting and taxation regulations that diverge from those imposed on non-financial sectors. Likewise, utility sectors frequently operate within heavily regulated frameworks where tax incentives and subsidies are commonplace, potentially distorting the connection between tax strategies and environmental or ESG considerations performance. This leaves the study with 1004 final observations after the sample selection criteria above. The sample selection procedure is reported in Table 1. Following the approach in [59], to mitigate the impact of outliers, we standardised all continuous variables by trimming at the 1% and 99% levels, ensuring that extreme values do not unduly influence our results.

**Table 1.** The table provides an overview of the sampling distribution in our study. Our sample covers 1004 firm-year observations from FTSE All Shares UK companies from 2014-2022.

| **Criteria** | **Number of Firm-Years** |
|---|---|
| Refinitiv Eikon | 1243 |
| Less: | |
| Utility Firms (SIC codes 6000-6999) | 44 |
| Financial Firms (SIC codes 4000-4999) | 186 |
| Others (SIC codes between 4000-4999 and 6000-6999) | 9 |
| Final sample | 1004 |

*3.2. Variable Measurement*

3.2.1. Dependent Variable

In our study, tax avoidance is estimated using the measure of total book-tax difference (*TBTD*) values following [60,61]. An increase in a firm's *TBTD* value indicates greater tax avoidance [62]. The method for calculating *TBTD* is as follows:

$$TBTD = \frac{TXDI_t + \left[[STR_t - ETR_t] \times PI_t\right]}{AT_{t-1}} \quad (1)$$

where $TXDI_t$ is the deferred tax expense in year *t*, $STR_t$is the highest UK corporate statutory tax rate in year *t*, $ETR_t$is the income tax expense divided by the pretax income in year *t*, $PI_t$ is the pretax income on year *t*, and $AT_{t-1}$ is a total asset in year *t – 1.*

#### 3.2.2. Independent Variable

Our study uses environmental performance (*EPILLAR*) as the independent variable to assess how much a company prioritises environmental factors within the environmental, social, and governance (*ESG*) framework and how this may influence the tax avoidance behaviour. For our research, environmental performance is defined as the environmental pillar score based on the work of Albitar et al. [56].

#### 3.2.3. Control Variables

To address various factors influencing tax avoidance, the following control variables are included in our study: return on assets (*ROA*), which is measured as income after taxes for the fiscal period divided by the average total assets [31,52,56]; size of the firm (*SIZE*), which is measured by the natural logarithm of total assets [63]; free cash flow (*FCF*), which is measured as cash flow from operations divided by total sales [56]; leverage (*LEV*), which is measured by total debt divided by total assets [2,31,52,64]; market-to-book ratio (*MTB*), which is measured by company market capitalisation divided by book value capitalisation [52,62]; asset/income ratio (*AIN*), which is measured by total assets divided by net income before taxes [31]; sales growth (*SG*), which is measured as the sales from year $t$ minus the sales from $t - 1$ divided by the sales from $t - 1$ [65]; liquidity ratio (*LIQ*), which is measured as current assets divided by current liabilities [56]; firm's age (*AGE*), which is measured by the natural logarithm of year $t$ minus the date of incorporation plus 1 [31]; big 4 (*BIG4*) serves as a dummy variable to denote whether the auditor is affiliate with one of the BIG 4 auditor firms (1) or not (0) [52,66,67]; board size (*BODSIZE*), which is measured by the natural logarithm of the number of board members [56]; board independence (*BODIND*), which is measured by the proportion of independent directors on the board [56,68]. We have also included a control variable in the form of financial transparency (*FFIN*), which is measured by industry and year. Financial transparency is calculated as the absolute value of a firm's scaled accruals, averaged over the prior three years, following previous research conducted in [69–72].

### *3.3. Regression Model*

Following Elamer et al. [74], our study employs the regression model technique to investigate the relationship between environmental performance and tax avoidance. Our study introduces (i) a model, (1), that does not incorporate controls or fixed effects (see Equation (2)), (ii) a model, (2), that includes controls but excludes fixed effects (see Equation (3)), and (iii) a model, (3), designed to present comprehensive results (see Equation (4)). Here are the models:

$$TBTD_{it} = \beta_0^{(1)} + \beta_1^{(1)} EPILLAR_{it} + \varepsilon_{it}^{(1)} \quad (2)$$

$$\begin{aligned} TBTD_{it} = \beta_0^{(2)} &+ \beta_1^{(2)} EPILLAR_{it} + \beta_2^{(2)} ROA_{it} + \beta_3^{(2)} SIZE_{it} + \beta_4^{(2)} FCF_{it} + \beta_5^{(2)} LEV_{it} \\ &+ \beta_6^{(2)} MTB_{it} + \beta_7^{(2)} AIN_{it} + \beta_8^{(2)} SG_{it} + \beta_9^{(2)} LIQ_{it} + \beta_{10}^{(2)} AGE_{it} \\ &+ \beta_{11}^{(2)} BIG4_{it} + \beta_{12}^{(2)} BODSIZE_{it} + \beta_{13}^{(2)} BODIND_{it} + \beta_{14}^{(2)} FFIN_{it} + \varepsilon_{it}^{(2)} \end{aligned} \quad (3)$$

$$\begin{aligned} TBTD_{it} = \beta_0^{(3)} &+ \beta_1^{(3)} EPILLAR_{it} + \beta_2^{(3)} ROA_{it} + \beta_3^{(3)} SIZE_{it} + \beta_4^{(3)} FCF_{it} + \beta_5^{(3)} LEV_{it} \\ &+ \beta_6^{(3)} MTB_{it} + \beta_7^{(3)} AIN_{it} + \beta_8^{(3)} SG_{it} + \beta_9^{(3)} LIQ_{it} + \beta_{10}^{(3)} AGE_{it} \\ &+ \beta_{11}^{(3)} BIG4_{it} + \beta_{12}^{(3)} BODSIZE_{it} + \beta_{13}^{(3)} BODIND_{it} + \beta_{14}^{(3)} FFIN_{it} \\ &+ Industry\ Sector\ Fixed\ Effect + Year\ Fixed\ Effect + \varepsilon_{it}^{(3)} \end{aligned} \quad (4)$$

In models (2) – (4), *TBTD* represents a tax avoidance measure, as explained in Section 3.2.1. *EPILLAR* is the index of environmental performance, as described in Section 3.2.2. All control variables are explained in Section 3.2.3. *Industry Sector* (based on Fama–French 12 industry classification) *Effects* and *Year Fixed Effects* are also controlled.

## 4. EMPIRICAL RESULTS

### *4.1. Descriptive Statistics*

#### 4.1.1. Sample Distribution by Industry and Year

Panel A in Table 2 illustrates the sampling distribution over the years ($l$), while Panel B presents the sampling distribution categorised by industry sectors ($k$) according to the Fama–French 12 industry sector classification. N.B. Yearly observations ($l$ = 1 to 9) indicate a lower total of 80 in 2014 and a peak of 141 in 2022. See above, in Section 3.1, why $k \leq 10$.

**Table 2.** Panel A: Sample distribution by year. Panel B: Sample distribution by industry sector.

| | | A: | | |
|---|---|---|---|---|
| ***l*** | **Year** | **Freq.** | **Percent** | **Cum.** |
| 1 | 2014 | 80 | 7.97 | 7.97 |
| 2 | 2015 | 96 | 9.56 | 17.53 |
| 3 | 2016 | 97 | 9.66 | 27.19 |
| 4 | 2017 | 102 | 10.16 | 37.35 |
| 5 | 2018 | 106 | 10.56 | 47.91 |
| 6 | 2019 | 116 | 11.55 | 59.46 |
| 7 | 2020 | 129 | 12.85 | 72.31 |
| 8 | 2021 | 137 | 13.65 | 85.96 |
| 9 | 2022 | 141 | 14.04 | 100.00 |
| | Total | 1004 | 100.00 | |

**B:**

| *k* | **Industry Sector** | **Freq.** | **Percent** | **Cum.** |
|---|---|---|---|---|
| 1 | Consumer Nondurables | 78 | 7.77 | 7.77 |
| 2 | Consumer Durables | 7 | 0.70 | 8.47 |
| 3 | Manufacturing | 149 | 14.84 | 23.31 |
| 4 | Oil, Gas, and Coal Extraction and Products | 48 | 4.78 | 28.09 |
| 5 | Chemicals and Allied Products | 23 | 2.29 | 30.38 |
| 6 | Business Equipment | 104 | 10.36 | 40.74 |
| 7 | Telephone and Television Transmission | 16 | 1.59 | 42.33 |
| 8 | Whole sales, Retail, and Some Services | 234 | 23.31 | 65.64 |
| 9 | Healthcare, Medical Equipment, and Drugs | 34 | 3.39 | 69.02 |
| 10 | Other | 311 | 30.98 | 100.00 |
| | Total | 1004 | 100.00 | |

4.1.2. Correlation Matrix

Table 3 presents the correlation matrix for all variables used in our analysis. First, this study documents a positive and significant association between environmental performance (*EPILLAR*) and tax avoidance (*TBTD*), with a correlation coefficient of 0.100. The findings provide preliminary evidence supporting our main hypothesis. While the relationship is statistically significant at the 1% level, it is relatively weak. A statistically significant positive relationship is observed between return on assets (*ROA*) and *TBTD*, with a coefficient of 0.174 at the 1% significance level. This indicates that firms with higher profitability are more likely to engage in tax avoidance activities. Similarly, the asset-to-income ratio (*AIN*) demonstrates a weak positive relationship with tax avoidance (coefficient = 0.096), also statistically significant at the 1% level, suggesting that firms with larger assets are more inclined toward tax avoidance. Lastly, the company's age exhibits a weak positive correlation with tax avoidance (coefficient = 0.054), significant at the 10% level, indicating that older firms are marginally more likely to engage in tax avoidance.

**Table 3.** Correlation matrix; *m* refers to the order of the variables in the model, starting from the main independent variable to the control variables.

| *m* | Variables | | 1 | 2 | 3 | 4 | 5 | 6 | 7 | 8 | 9 | 10 | 11 | 12 | 13 | 14 |
|---|---|---|---|---|---|---|---|---|---|---|---|---|---|---|---|---|
| | *TBTD* | 1.000 | | | | | | | | | | | | | | |
| 1 | *EPILLAR* | 0.100 *** | 1.000 | | | | | | | | | | | | | |
| 2 | *ROA* | 0.174 *** | −0.115 *** | 1.000 | | | | | | | | | | | | |
| 3 | *SIZE* | −0.046 | 0.596 *** | −0.148 *** | 1.000 | | | | | | | | | | | |
| 4 | *FCF* | −0.076 ** | −0.310 *** | 0.287 *** | −0.387 *** | 1.000 | | | | | | | | | | |
| 5 | *LEV* | 0.008 | 0.183 *** | −0.222 *** | 0.139 *** | −0.265 *** | 1.000 | | | | | | | | | |
| 6 | *MTB* | −0.062 ** | −0.206 *** | 0.607 *** | −0.323 *** | 0.471 *** | −0.064 ** | 1.000 | | | | | | | | |
| 7 | *AIN* | 0.096 *** | −0.122 *** | 0.035 | −0.200 *** | 0.107 *** | −0.052 * | 0.049 | 1.000 | | | | | | | |
| 8 | *SG* | 0.047 | −0.023 | 0.152 *** | −0.080 ** | 0.082 *** | 0.004 | 0.047 | 0.051 * | 1.000 | | | | | | |
| 9 | *LIQ* | −0.018 | −0.159 *** | 0.171 *** | −0.280 *** | 0.190 *** | −0.227 *** | 0.145 *** | 0.012 | 0.114 *** | 1.000 | | | | | |
| 10 | *AGE* | 0.054 * | 0.043 | −0.044 | 0.142 *** | −0.155 *** | −0.222*** | −0.220 *** | 0.022 | −0.097 *** | 0.024 | 1.000 | | | | |
| 11 | *BODSIZE* | −0.039 | 0.495 *** | −0.158 *** | 0.587 *** | −0.281 *** | 0.218 *** | −0.145 *** | −0.195 *** | −0.035 | −0.106 *** | −0.048 | 1.000 | | | |
| 12 | *BODIND* | −0.012 | 0.374 *** | −0.090 *** | 0.375 *** | −0.239 *** | 0.080 ** | −0.180 *** | −0.066 ** | −0.034 | −0.104 *** | 0.174 *** | 0.312 *** | 1.000 | | |
| 13 | *BIG4* | −0.058 * | −0.038 | 0.079 ** | −0.060 * | −0.003 | −0.009 | 0.104*** | 0.042 | −0.005 | −0.045 | 0.114 *** | −0.058 * | −0.051 * | 1.000 | |
| 14 | *FFIN* | 0.005 | 0.008 | −0.016 | 0.009 | 0.048 | −0.039 | 0.005 | −0.005 | −0.015 | −0.064 ** | −0.058 * | 0.033 | −0.025 | 0.025 | 1.000 |

*** $p < 0.01$, ** $p < 0.05$, * $p < 0.1$.

4.1.3. Summary Statistics and Variance Inflation Factor Test

Table 4 presents the sample distribution for all variables used in the analysis (Panel A) and the variance inflation factor (*VIF*) test to check for multicollinearity. On average, the environment performance score is 0.484, suggesting a percentage value of 48.4% on their environmental performance. While some companies received scores as low as 0, others neared the upper limit of 0.958, indicating a wide range of environmental performance practices. Returns on assets (*ROA*s) have an average value of 0.065, showing that firms are generally profitable (positive *ROA*). The sample's average size indicates that most companies are large, with a mean value of 21.242 (natural logarithm), reflecting the characteristics of these firms. The mean free cash flow (*FCF*) is 0.084, with some firms having no free cash flow and others reporting values up to 0.460. Financial leverage (*LEV*) has an average value of 0.214, with some having high leverage (up to 0.770). The market-to-book (*MTB*) is observed to have a mean value of 2.014. The average sales growth (*SG*) is positive (mean value = 0.093), with some firms experiencing declines (negative values) and others achieving significant growth. Most firms maintain moderate liquidity (*LIQ*), but some have high liquidity levels (up to 8.112). Among the sampled firms, 0.649 or ≃65% use the service of *BIG4* firms, with an average board size (*BODSIZE*) ≃9 and board independence (i) of 0.617, i.e., ≃62%. *FFIN* indicates that 70% of firms in the dataset are not transparent. The results for *VIF* for the multicollinearity test (Panel B) confirm no signs of multicollinearity in the model, as indicated by *VIF* values < 10 and 1/*VIF* values > 0.10.

**Table 4.** Panel A. Summary Statistics. Panel B. Variance inflation factors (VIFs). The Sum value (in panel A) includes all observations collected from 2014 to 2022, featuring 567 companies, after data cleaning and winsorizing processes; *m* refers to the order of the variables in the models, starting from the main independent variable and ending with the control variables.

**A**

| *M* | | Sum | Mean | StDev | Min | Median | Max |
|---|---|---|---|---|---|---|---|
| | *TBTD* | 1004 | −0.007 | 0.018 | −0.092 | −0.005 | 0.068 |
| 1 | *EPILLAR* | 1004 | 0.484 | 0.233 | 0.000 | 0.463 | 0.958 |
| 2 | *ROA* | 1004 | 0.065 | 0.101 | −0.248 | 0.054 | 0.448 |
| 3 | *SIZE* | 1004 | 21.242 | 1.642 | 16.835 | 20.990 | 25.512 |
| 4 | *FCF* | 1004 | 0.084 | 0.087 | 0.000 | 0.060 | 0.460 |
| 5 | *LEV* | 1004 | 0.214 | 0.158 | 0.000 | 0.210 | 0.770 |
| 6 | *MTB* | 1004 | 2.014 | 2.202 | 0.235 | 1.311 | 12.875 |
| 7 | *AIN* | 1004 | 0.150 | 0.636 | −2.727 | 0.062 | 4.391 |
| 8 | *SG* | 1004 | 0.093 | 0.272 | −0.552 | 0.060 | 1.874 |
| 9 | *LIQ* | 1004 | 1.571 | 1.088 | 0.242 | 1.320 | 8.112 |
| 10 | *AGE* | 1004 | 3.115 | 1.008 | 0.000 | 3.040 | 4.930 |
| 11 | *BODSIZE* | 1004 | 8.840 | 2.266 | 3.000 | 8.000 | 17.000 |
| 12 | *BODIND* | 1004 | 0.617 | 0.138 | 0.000 | 0.625 | 0.867 |
| 13 | *BIG4* | 1004 | 0.649 | 0.477 | 0.000 | 1.000 | 1.000 |
| 14 | *FFIN* | 1004 | 0.700 | 0.458 | 0.000 | 1.000 | 1.000 |

**B**

| | *VIF* | *1/VIF* |
|---|---|---|
| *SIZE* | 2.310 | 0.434 |
| *MTB* | 2.140 | 0.467 |
| *ROA* | 1.790 | 0.559 |
| *BODSIZE* | 1.760 | 0.568 |
| *EPILLAR* | 1.710 | 0.584 |
| *FCF* | 1.540 | 0.647 |
| *LEV* | 1.290 | 0.774 |
| *BODIND* | 1.260 | 0.791 |
| *AGE* | 1.230 | 0.810 |
| *LIQ* | 1.190 | 0.843 |
| *SG* | 1.060 | 0.943 |
| *AIN* | 1.060 | 0.943 |
| *BIG4* | 1.050 | 0.953 |
| *FFIN* | 1.020 | 0.982 |
| *Mean VIF* | 1.460 | |

### *4.2. Environmental Performance and Tax Avoidance (Hypothesis 1)*

Table 5 presents the regression analysis of the relationship between environmental performance and tax avoidance. The results indicate a positive association between environmental performance (*EPILLAR*) and tax avoidance (*TBTD*). The results in column (2) provide the initial findings for this positive and significant association, even without control and fixed-effect variables (refer to Equation (2)).

In column (3) (refer to Equation (3)), *EPILLAR* exhibits a positive and significant value, with a coefficient of 0.160. Furthermore, *ROA* appears to have a positive and significant impact, indicating that higher profitability (*ROA*) is linked to increased tax avoidance (*TBTD*). *AIN* also shows a positive and significant relationship, suggesting that higher asset intensity contributes to greater tax avoidance. However, the company's size demonstrates a negative and significant correlation, implying that larger companies (in terms of revenue) have lower tax avoidance. Moreover, *FCF* is negatively significant, indicating that higher free cash flow is associated with lower tax avoidance. The same goes for *MTB,* which also shows a negative and significant relationship, implying that a higher market valuation relative to book value is linked to lower tax avoidance. Lastly, being audited by a Big Four firm exhibits negative and significant correlations, suggesting that undergoing an audit by a Big Four firm is associated with lower tax avoidance. The findings concerning the lower propensity to engage in tax avoidance among larger more established companies support the previous literature [27,73]. Larger companies typically possess greater resources to manage their operations and exhibit heightened concern regarding the potential reputational risks associated with tax non-compliance. The observed positive relationship between *ROA* and tax avoidance may initially seem unexpected but can be rationalised through the growth factor. Firms may need to reinvest their profits to sustain high rates of return, as evidenced by research on the tax avoidance behaviours of high-growth firms in the Chinese context [36].

In column (4) (refer to Equation (4)), the fixed-effects model adjusts for industry-specific factors, yielding more robust results. *EPILLAR* consistently shows a positive and significant relationship with tax avoidance across all models. This result is statistically significant, with a coefficient of 0.154 (*t*-test = 4.958). Variables such as *ROA*, *SIZE*, *MTB*, *AIN*, *SG*, *LIQ*, and *AGE* consistently demonstrate significance, underscoring their importance in explaining tax avoidance (*TBTD*). The fixed-effects model with clustered standard errors provides the most reliable estimates by controlling for industry-specific effects and addressing potential heteroskedasticity. The findings suggest that companies engaging more intensively in environmental activities (or high environmental performance) are more likely to practice tax avoidance. See our H1.

**Table 5.** Baseline regression results of models (2) – (4).

| *m* | | **Dependent Variable:** ***TBTD*** | | |
|---|---|---|---|---|
| | | **(2)** | **(3)** | **(4)** |
| 1 | *EPILLAR* | 0.078 *** (3.192) | 0.160 *** (5.328) | 0.154 *** (4.958) |
| 2 | *ROA* | | 0.067 *** (9.354) | 0.067 *** (3.901) |
| 3 | *SIZE* | | −0.003 *** (−5.147) | −0.002 *** (−3.514) |
| 4 | *FCF* | | −0.014 * (−1.851) | −0.014 (−0.844) |
| 5 | *LEV* | | 0.006 * (1.653) | 0.007 (0.931) |
| 6 | *MTB* | | −0.002 *** (−6.164) | −0.002 *** (−4.352) |
| 7 | *AIN* | | 0.002 *** (2.838) | 0.002 ** (2.975) |
| 8 | *SG* | | 0.000 (0.037) | 0.003 ** (2.514) |
| 9 | *LIQ* | | −0.001 * (−1.806) | −0.001 * (−1.965) |
| 10 | *AGE* | | 0.001 * (1.740) | 0.001 ** (2.299) |
| 11 | *BODSIZE* | | 0.000 (0.219) | −0.000 (−0.280) |
| 12 | *BODIND* | | −0.007 (−1.642) | −0.007 (−1.143) |
| 13 | *BIG4* | | −0.003 *** (−2.673) | −0.003 (−1.792) |
| 14 | *FFIN* | | 0.001 (0.618) | 0.001 (0.876) |
| | Constant | −0.011 *** (−8.278) | 0.043 *** (4.360) | 0.027 ** (2.705) |
| | Number of Observations | 1004 | 1004 | 1004 |
| | Adjusted R-squared | 0.009 | 0.123 | 0.158 |
| | Year Fixed Effect | NO | NO | YES |
| | Industry Sector Fixed Effect | NO | NO | YES |
| | Cluster by Industry | NO | NO | YES |

Table 5 presents the baseline regression results. The dependent variable is tax avoidance measured by *TBTD*, and the main independent variable is environmental performance measured by *EPILLAR*. Robust t-test results are reported in parentheses next to the coefficients. Standard errors are corrected at the firm level. ***, **, and * denote significance at the 1%, 5%, and 10% levels, respectively. Variable definitions are provided in Appendix A.

### *4.3. The Moderating Role of Financial Constraints on the Relationship Between Environmental Performance and Tax Avoidance (Hypothesis 2)*

Our second hypothesis (H2) posits that the impact of environmental performance on corporate tax avoidance varies depending on the degree of financial constraints faced by firms. Specifically, firms experiencing significant financial constraints are more likely to engage in aggressive tax planning compared to their less constrained counterparts [23,24]. Financially constrained firms may view tax evasion as a practical strategy to conserve resources, particularly when they are concurrently investing in environmental initiatives [25]. Our study follows the literature to assess the moderating effect of financial constraints on the documented relationship between environmental performance and tax avoidance [24,58,75,76]. It employs three measures to proxy for financial constraints. Initially, our study considers sales as in [24], WW score [58], and KZ score [75–77]. Following the approach in [78], our study utilises an indicator variable, which takes a value of 1 as a high financial constraint and 0 as a low financial constraint for WW score and for KZ score to examine its moderating effect on the documented relationship. The firm's sales are used to measure the level of the firm's financial constraints, in line with the methodology of [76]. We define a reverse scale of the variable "high sales" = 1 if the firm's sales exceed the median sample value, and zero otherwise. High sales indicate that the company faces low financial constraints, whereas low sales suggest that the company faces high financial constraints [24]. As reported in columns labelled (1) – (2) in Table 6, it is apparent that companies with low sales (indicating high financial constraints) tend to practice tax avoidance through environmental performance.

Although companies with high sales (indicating low financial constraints) also engage in tax avoidance, there are discernible differences (0.131) in the impact between companies with low financial constraints (0.088). Following the prior literature, our study utilises the WW score as the second proxy for financial constraints [57,58,79–81]. A higher WW score indicates more stringent financial constraints for the company [57]. The findings in columns (3) – (4) suggest that companies with a high WW score are more inclined to practice tax avoidance. This observation is consistent with the results in columns (1) – (2), which suggest that financially constrained firm tend to engage in tax avoidance more than unconstrained companies.

Lastly, our study utilises the KZ score to investigate the impact of financial constraints on the relationship between environmental intensity and tax avoidance. The KZ score, initially developed by [82], incorporates five accounting variables—cash flow, Tobin's Q, leverage, dividends, and cash. Baker et al. [77] later modified this formula by removing the Q variable based on the concept of [83] while still achieving similar results. Our research uses the modified formula of [77].

Our findings, see columns (5) – (6), Table 6, indicate that high financial constraints, as indicated by a high KZ score, influence companies to engage in tax avoidance (0.178). However, it is also possible for companies with low financial constraints to engage in tax avoidance (0.133) with a 10% significance level.

Nonetheless, the distinction between firms facing high and low financial constraints, as assessed by the KZ score method, does not indicate a significant difference. This differs from the outcomes associated with sales and WW scores. The permutation test results for varying coefficients related to the KZ score exceed the 10% significance level ($p$-value = 0.300).

**Table 6.** The moderating role of financial constraints on the relationship between environmental performance and tax avoidance.

| | **Dependent Variable:** ***TBTD*** | | | | | |
|---|---|---|---|---|---|---|
| **Financial Constraints:** | **High Sales** | **Low Sales** | **High WW Score** | **Low WW Score** | **High KZ** | **Low KZ** |
| | **(1)** | **(2)** | **(3)** | **(4)** | **(5)** | **(6)** |
| *EPILLAR* | 0.088 *** (3.937) | **0.219** *** (3.569) | **0.246** *** (3.469) | 0.074 ** (3.325) | **0.178** * (1.954) | 0.133 * (2.027) |
| Controls | YES | YES | YES | YES | YES | YES |
| Constant | 0.008 (0.364) | 0.078 ** (3.185) | 0.049 * (2.210) | 0.024 (0.988) | 0.029 (1.781) | 0.019 ** (2.927) |
| Number of Observations | 501 | 503 | 502 | 502 | 502 | 502 |
| Adjusted R-squared | 0.114 | 0.225 | 0.159 | 0.314 | 0.145 | 0.223 |
| Year Fixed Effect | YES | YES | YES | YES | YES | YES |
| Industry Sector Fixed Effect | YES | YES | YES | YES | YES | YES |
| Cluster by Industry | YES | YES | YES | YES | YES | YES |
| Permutation tests for coeff. | $p$-value < 0.000 | | $p$-value < 0.000 | | $p$-value = 0.300 | |

Table 6 presents the regression results for the moderating effect of financial constraints on the relationship between environmental performance and tax avoidance. The dependent variable is tax avoidance measured by *TBTD*, and the main independent variable is environmental performance measured by *EPILLAR*. The Sales, WW score, and KZ index values are the proxies used for financial constraints. Robustness t-test is reported in parentheses. Standard errors are corrected at the firm level. ***, **, and * denote significance at the 1%, 5%, and 10% levels, respectively. Variable definitions are provided in Appendix A. The table illustrates that companies with high financial constraints and those with low financial constraints have similar opportunities for tax avoidance through their environmental performance. However, companies facing high financial constraints demonstrate a greater degree of tax avoidance than their counterparts with low financial constraints, as indicated by the values highlighted in bold face.

### *4.4. Endogeneity and Robustness Tests*

#### 4.4.1. Introduction to Entropy Considerations

Entropy balancing is a concept utilised to ascertain endogeneity by segmenting data into two distinct groups and establishing a benchmark to ensure that these groups possess perfectly aligned features and characteristics [84]. Our paper examines the environmental pillar as a benchmark for assessing companies' engagement in environmental activities. However, it cannot be presumed that companies exhibiting high levels of environmental engagement will experience identical tax avoidance effects to those with lower levels. Therefore, executing an entropy balancing is essential for mitigating bias in the analysis. Below is the formula used to measure the (information) entropy, as per [85]:

$$Entropy = -\sum_{x=1}^{M} p_x \, ln(p_x) \qquad (5)$$

where $p_x$ represents the probability of a given outcome, $M$ denotes the total number of possible outcomes, and $ln(p_x)$ signifies the natural logarithm of the probability. Our research utilises *EPILLAR* to categorise the data into two distinct groups. The following illustrates the workings of entropy within this context. The function $p_x$ can be defined as a probability, fraction, or ratio:

$$p_x = \frac{N_x}{\sum_{x=1}^{M} N_x} \qquad (6)$$

where $M$ represent the number of bins; we use 10 bins in this analysis. The variable $N_x$ denotes the value corresponding to the $x$-th bin, with $\sum_{x=1}^{M} N_x$ which is defined as $N$, the total number of observations = 1004 (see Table 3). For illustrative purposes, a histogram depicting the distribution of *EPILLAR* is provided (refer to Figure 1).

**Table 7.** Entropy balancing and propensity score Analysis. Panel A: The calculation of *Entropy* for *EPILLAR*. Panel B: Univariate comparison of means between treatment and control groups before and after balancing—entropy balancing. Panel C: Univariate comparison of means between treatment and control groups—PSM is propensity score. Panel D: Regression estimates.

**A**

| $x$ | $N_{EPILLAR}$ | $\frac{N_x}{N}$ | $ln(p_x)$ | $Entropy_{EPILLAR}$ |
|---|---|---|---|---|
| 0 | | | | |
| 1 | 39 | 0.038845 | −3.248186 | 0.126175 |
| 2 | 81 | 0.080677 | −2.517298 | 0.203089 |
| 3 | 132 | 0.131474 | −2.028945 | 0.266754 |
| 4 | 123 | 0.12251 | −2.099563 | 0.257217 |
| 5 | 183 | 0.182271 | −1.702261 | 0.310273 |
| 6 | 130 | 0.129482 | −2.044213 | 0.264689 |
| 7 | 89 | 0.088645 | −2.423111 | 0.214798 |
| 8 | 94 | 0.093625 | −2.368453 | 0.221748 |
| 9 | 102 | 0.101594 | −2.286774 | 0.232322 |
| 10 | 31 | 0.030876 | −3.47776 | 0.107381 |
| | | | Sum | 2.204444 |

**B**

| Variable | High Intensity | Low Intensity | High Intensity | Low Intensity |
|---|---|---|---|---|
| | Before Balancing | | After Balancing | |
| *ROA* | 0.057 | 0.074 | 0.057 | 0.057 |
| *SIZE* | 21.940 | 20.540 | 21.940 | 21.940 |
| *FCF* | 0.066 | 0.103 | 0.066 | 0.066 |
| *LEV* | 0.238 | 0.191 | 0.238 | 0.238 |
| *MTB* | 1.665 | 2.362 | 1.665 | 1.665 |
| *AIN* | 0.091 | 0.209 | 0.091 | 0.091 |
| *SG* | 0.097 | 0.089 | 0.097 | 0.097 |
| *LIQ* | 1.478 | 1.664 | 1.478 | 1.478 |
| *AGE* | 3.162 | 3.069 | 3.162 | 3.162 |
| *BODSIZE* | 9.733 | 7.946 | 9.733 | 9.733 |
| *BODIND* | 0.656 | 0.579 | 0.656 | 0.656 |
| *BIG4* | 0.618 | 0.681 | 0.618 | 0.618 |
| *FFIN* | 0.703 | 0.697 | 0.703 | 0.703 |

**C**

| Variable | High Intensity | Low Intensity | Difference in Mean | |
|---|---|---|---|---|
| | (Treated) | (Control) | Diff | *t*-value |
| *ROA* | 0.060 | 0.055 | −0.005 | −0.497 |
| *SIZE* | 21.028 | 21.009 | −0.019 | −0.154 |
| *FCF* | 0.088 | 0.082 | −0.006 | −0.811 |
| *LEV* | 0.226 | 0.227 | 0.001 | 0.092 |
| *MTB* | 1.931 | 1.961 | 0.030 | 0.164 |
| *AIN* | 0.138 | 0.125 | −0.014 | −0.232 |
| *SG* | 0.097 | 0.096 | −0.001 | −0.016 |
| *LIQ* | 1.623 | 1.622 | −0.001 | −0.015 |
| *AGE* | 3.214 | 3.166 | −0.048 | −0.524 |
| *BODSIZE* | 8.700 | 8.708 | 0.008 | 0.046 |
| *BODIND* | 0.616 | 0.612 | −0.004 | −0.337 |
| *BIG4* | 0.648 | 0.657 | 0.009 | 0.194 |
| *FFIN* | 0.700 | 0.670 | −0.030 | −0.697 |

**D**

| VARIABLES | Entropy Balancing | PSM |
|---|---|---|
| *EPILLAR* | 0.175 *** (4.608) | 0.127 ** (2.685) |
| Controls | YES | YES |
| Constant | 0.027 * (2.126) | 0.013 (0.637) |
| Observations | 1004 | 392 |
| Adjusted R-squared | 0.165 | 0.170 |
| Year Fixed Effect | YES | YES |
| Industry Sector Fixed Effect | YES | YES |
| Cluster by Industry | YES | YES |

Panel 7A presents the calculation of the *EPILLAR* distribution entropy. Panel 7B proposes the entropy balancing results of the univariate comparison of means between treatment and control groups before and after balancing, distinguishing high vs. low intensity. Panel 7C reports the PSM results of the univariate comparison of means between treatment and control groups. Panel 7D presents the regression results after propensity score matching analysis (PSM) and entropy balancing. The dependent variable is tax avoidance measured by *TBTD*; the main independent variable is the environmental performance measured by *EPILLAR*. Robustness t-statistics are reported in parentheses next to the coefficients. Standard errors are corrected at the firm level. ***, **, and * denote significance at the 1%, 5%, and 10% levels, respectively. Variable definitions are provided in Appendix A.

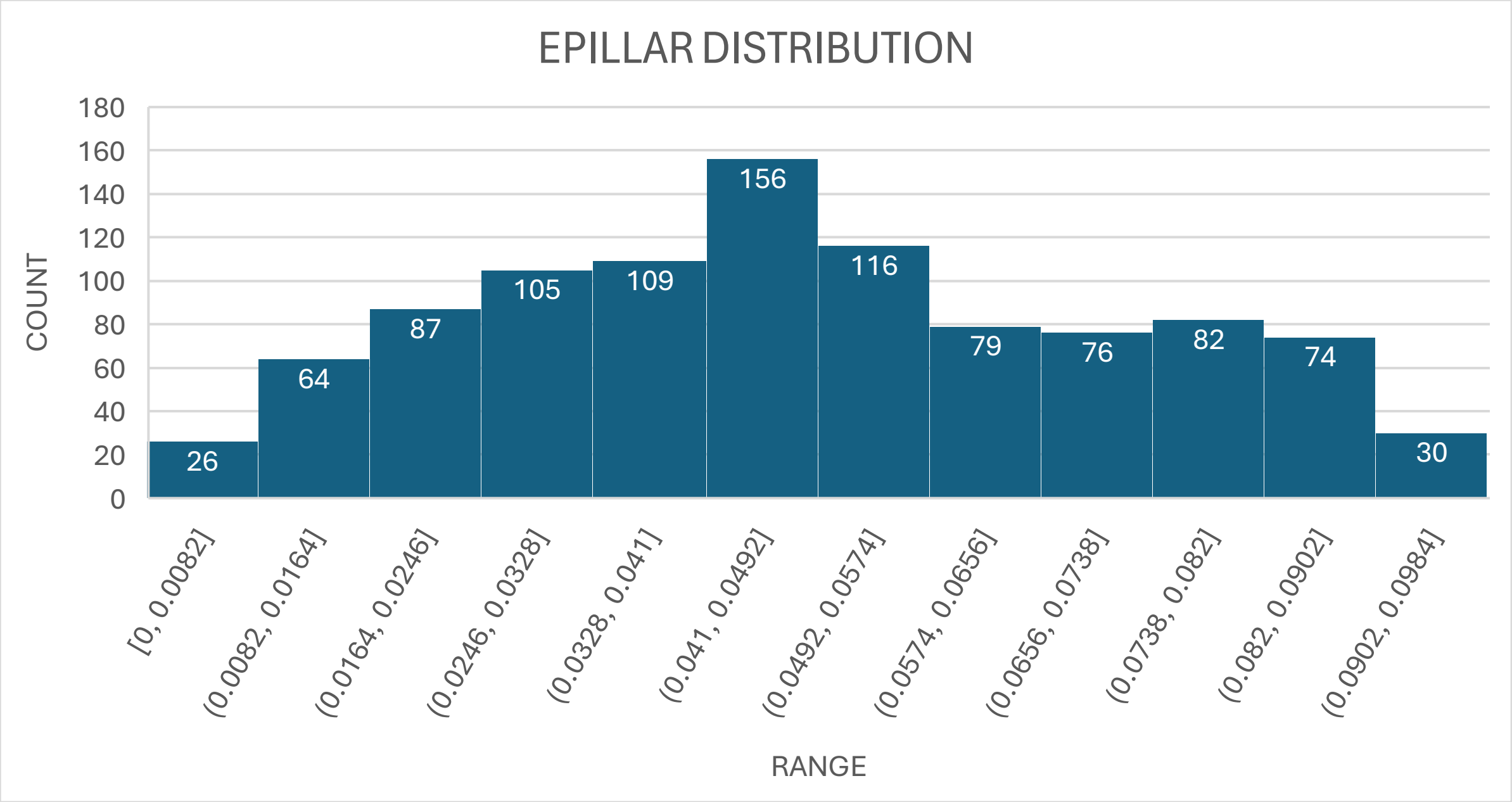

**Figure 1.** The distribution of *EPILLAR* is characterised by a data range extending from 0 to 0.1. It is evident that each bin—we here arbitrarily utilise ten bins to delineate the distribution of *EPILLAR* —exhibits a value (denoted as $N_x$). Furthermore, the cumulative value of all bins corresponds to the total number of observations ($N$ = 1004).

As illustrated in Figure 1, the first bin ($N_1$) value is 39, and the total observation count (the sum of the value on each bin) ($N$) is 1004. Consequently, we can derive that for the first bin, $p_x$ is 0.038845. The natural logarithm of 0.038845 is equal to −3.24819. In a similar manner, one can compute the probability associated with each bin, and the data entropy is represented by the cumulative value of the probabilities from bin 1 to the last bin 10. The entropy of a uniform distribution, defined as when $p_x$= 1/$N$, is equal to ln($N$). This represents the maximum entropy value, corresponding to a state of maximum uncertainty, where all outcomes are equally likely, indicating a lack of structure or predictability. In our case, ln(1004) ≃ 6.91175; defining $N_{EPILLAR}$ as the number distribution of each *EPILLAR* per bar and $Entropy_{EPILLAR}$ as the calculation of *Entropy* for each bar, one obtains the results in Table 7. All distributions (see column 3 on Table 7 Panel A) deviate significantly from uniformity, indicating the necessity for a balancing process; see Figure 2.

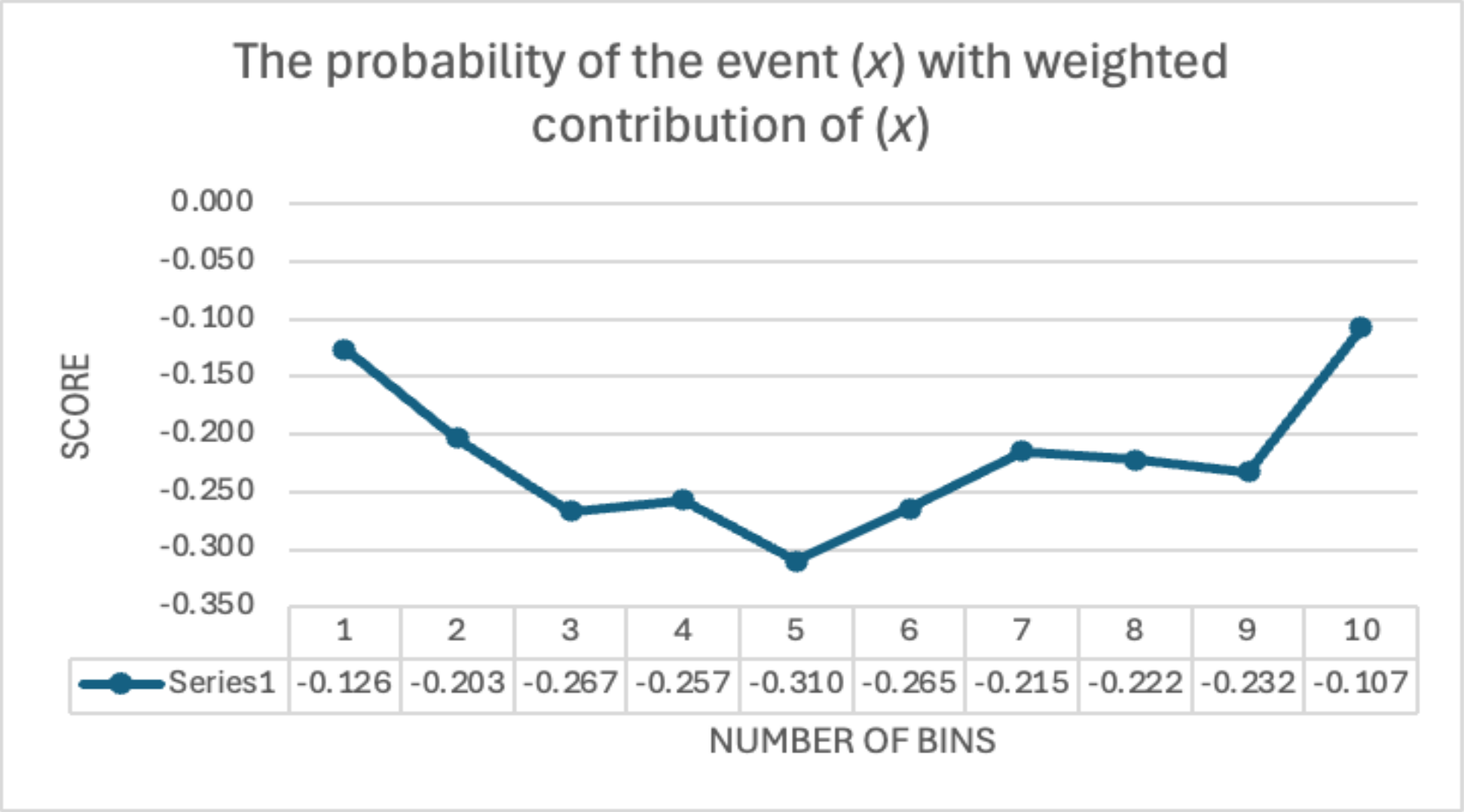

**Figure 2.** The distribution of ($p_x$)(ln($p_x$)). The variable denoted by $x$ represents the number of bins, ranging from 1 to 10. It is pertinent to emphasise that there exists no "entropy" within a bin designated as "$x$"; rather, the "entropy" is quantified as the (negative) summation of $p_x$ ln($p_x$).

4.4.2. Entropy Balancing and Propensity Score Matching

Thereafter, our study first employs a propensity score matching (PSM) analysis to address this concern. The sample is divided into two groups based on the median environmental performance (*EPILLAR*) score, and logistic regression is then utilised to estimate propensity scores for predicting high environmental performance. All controls and fixed effects were included in the regression. Our study opts for one-to-one matching without replacement and sets a maximum calliper distance of 0.005, following [86]. Our study considers that firms with higher environmental performance levels may differ systematically from those with lower levels, potentially introducing bias into the estimates.

A significant drawback of the propensity score matching (PSM) method is its tendency to reduce sample sizes significantly. To overcome this challenge, our study instead applies the entropy balancing method suggested by [84] as an alternative matching strategy. Like PSM, entropy balancing seeks to achieve a covariate balance between treatment and control groups but does so while preserving a larger number of observations. From Panel B of Table 7, it is evident that there are significant differences in the mean values of variables between the treatment and control groups before balancing. This disparity could result in biased treatment effect estimates.

However, after balancing, the means of variables in the control group closely match those in the treatment group, signalling successful balancing. The results are reported in Panel 7B–D.

Notice that the sample size is reduced to 392 observations from the original 1004 due to the score matching, as pointed out in Panel D of Table 7. Notably, even after the matching process, the regression results remained significant, particularly for environmental performance (*EPILLAR*) and tax avoidance (*TBTD*). Considering the t-test value of the difference between the high- and low-environmental performance groups below 1.67 (see Panel C of Table 7), it can be inferred that the gap is modest, which aligns well with the model's suitability. As reported in Panel D of Table 7, these regression estimates show consistent results to the baseline findings after the PSM. The results suggest no endogeneity issues based on the entropy balancing technique and propensity score matching.

4.4.3. Heckman Model

In addition to the earlier endogeneity test, our study employs Heckman's two-stage test to address potential sample selection bias. In the first stage, a probit model is used to identify factors influencing the likelihood of being classified as high environmental performance. Our study uses the industry average (*IND_AVG*) as an instrument in the analysis. Utilising the industry average effectively captures all forms of variation within the sector. This approach is essential, as economic conditions and market dynamics are typically consistent within the same type of industry. The results (see Table 8) show that the industry average has a positive and highly significant coefficient of 0.439 (*t*-test = 4.592), indicating that a higher *IND_AVG* increases the likelihood of being classified as high environmental performance.

In the second stage, the *TBTD* is regressed on the predictor variables, including the inverse Mills ratio to correct for selection bias. This model includes significant coefficients, including *EPILLAR*, with a positive and significant coefficient of 0.154 (*t*-test = 4.982). This indicates that higher *EPILLAR* is associated with a higher *TBTD* value (see Table 8).

**Table 8.** Heckman sample selection bias test.

| | Dependent Variable: *TBTD* |
| --- | --- |

| | First Stage | Second Stage |
|---|---|---|
| *IND_AVG* | 0.439 *** (4.592) | - |
| *EPILLAR* | - | 0.154 *** (4.892) |
| Controls | YES | YES |
| Constant | −10.795 *** (−6.840) | 0.025 (0.837) |
| Observations | 1004 | 1004 |
| Year Fixed Effect | YES | YES |
| Industry Sector Fixed Effect | YES | YES |
| Cluster by Industry | YES | YES |
| Adjusted R-squared | 0.261 | 0.157 |

Table 8 presents the regression results after Heckman's sample selection bias test. In the first stage, the industry average is use as an instrument with environmental performance indicators being the dependent variable. In the second stage, the dependent variable is tax avoidance measured by *TBTD*, and the main independent variable is environmental performance measured by *EPILLAR*. Robustness t-statistics are reported in parentheses next to the coefficients. Standard errors are corrected at the firm level. *** denotes significance at the 1% level. Variable definitions are provided in Appendix A.

4.4.4. Instrumental Variable Approach

Furthermore, our study employs an instrumental variable (IV) regression analysis to address potential endogeneity issues further. The results are reported in Table 9. In the first stage, the coefficient for *IND_AVG* is significant, with a coefficient of 0.007 (*t*-test = 5.580), indicating statistical significance and a valid instrument to predict environmental performance (*EPILLAR*). The F-statistic for the instrument is 20.950, surpassing the critical value of 8.960, signifying the strength of *IND_AVG* as an instrument. In the second stage, the effect of the endogenous variable (*EIN*) on the dependent variable (*TBTD*) is estimated while controlling for other covariates and using *IND_AVG* as an instrument. The coefficient for *EPILLAR* is 0.513 (*t*-test = 1.820), showing marginal significance. The analysis highlights some significant findings, indicating that the conclusion drawn from the study remains robust after the above endogeneity tests.

**Table 9.** Instrument variables technique.

| | Dependent Variable: *TBTD* | |
|---|---|---|
| | **First Stage** | **Second Stage** |
| *IND_AVG* | 0.007 *** (5.58) | - |
| *EPILLAR* | - | 0.513 * (1.820) |
| Controls | YES | YES |
| Observations | 1004 | 1004 |
| Year Fixed Effect | YES | YES |
| Industry Sector Fixed Effect | YES | YES |
| Cluster by Industry | YES | YES |
| Cragg-Donald Wald F statistics | 20.950 | 20.951 |
| 15% Maximal IV size | 8.960 | - |
| Kleibergen–Paap rk F statistics | 29.120 | 29.117 |

Table 9 reports the regression estimate after using the instrumental variable approach. In the first stage, the study employs the industry average as an instrument with environmental performance indicators being the dependent variable. In the second stage, the dependent variable is tax avoidance measured by *TBTD*, and the main independent variable is environmental performance measured by *EPILLAR*. Robustness t-statistics are reported in parentheses next to the coefficients. Standard errors are corrected at the firm level. ***, and * denote significance at the 1%, and 10% levels, respectively. Variable definitions are provided in Appendix A.

4.4.5. An Alternative Measure of Tax Avoidance

In prior research, such as that conducted by Hanlon and Heitzman [73], alternative tax avoidance measures are employed to ensure the robustness of our findings. Specifically, our study uses the effective tax rate (*ETR*) (see Equations (7) and (8)) and book-tax difference (*TBTD*) measures (see Equation (2)). It is crucial to underline that *ETR* and *TBTD* have opposing influences. When a firm practices tax avoidance, the TBTD value rises, while the *ETR* value declines [87,88]. Therefore, we employ "*Current_ETR*" and "*GAAP_ETR*" as alternative tax avoidance measures (according to [73], the current effective tax rate (*ETR*) can be computed by dividing the current income tax expense by the total pre-tax accounting income. In contrast, the *GAAP* effective tax rate (*ETR*) is calculated by dividing the total income tax expense by the total pre-tax accounting income). The following equations are presented:

$$\begin{aligned} Current_ETR_{it} &= \alpha_0^{(1)} + \alpha_1^{(1)} EPILLAR_{it} + \alpha_2^{(1)} ROA_{it} + \alpha_3^{(1)} SIZE_{it} + \alpha_4^{(1)} FCF_{it} + \alpha_5^{(1)} LEV_{it} \\ &+ \alpha_6^{(1)} MTB_{it} + \alpha_7^{(1)} AIN_{it} + \alpha_8^{(1)} SG_{it} + \alpha_9^{(1)} LIQ_{it} + \alpha_{10}^{(1)} AGE_{it} + \alpha_{11}^{(1)} BIG4_{it} \\ &+ \alpha_{12}^{(1)} BODSIZE_{it} + \alpha_{13}^{(1)} BODIND_{it} + \alpha_{14}^{(1)} FFIN_{it} + Industry\ Sector\ Fixed\ Effect \\ &+ Year\ Fixed\ Effect + \eta'^{(1)}_{it} \end{aligned} \quad (7)$$

$$\begin{aligned} GAAP_ETR_{it} &= \alpha_0^{(2)} + \alpha_1^{(2)} EPILLAR_{it} + \alpha_2^{(2)} ROA_{it} + \alpha_3^{(2)} SIZE_{it} + \alpha_4^{(2)} FCF_{it} + \alpha_5^{(2)} LEV_{it} \\ &+ \alpha_6^{(2)} MTB_{it} + \alpha_7^{(2)} AIN_{it} + \alpha_8^{(2)} SG_{it} + \alpha_9^{(2)} LIQ_{it} + \alpha_{10}^{(2)} AGE_{it} + \alpha_{11}^{(2)} BIG4_{it} \\ &+ \alpha_{12}^{(2)} BODSIZE_{it} + \alpha_{13}^{(2)} BODIND_{it} + \alpha_{14}^{(2)} FFIN_{it} + Industry\ Sector\ Fixed\ Effect \\ &+ Year\ Fixed\ Effect + \eta'^{(2)}_{it} \end{aligned} \quad (8)$$

The regression results are reported in Table 10. The regression analysis reveals a link between a company's environmental performance and tax avoidance activities. The findings indicate that a 1-point rise in environmental performance corresponds to a 0.020-point decline in the current effective tax rate (*ETR*), suggesting increased tax avoidance alongside a 0.015-point drop in the *GAAP_ETR*. Nevertheless, the analysis underscores that the *TBTD* variable is more suitable for capturing tax avoidance due to its stronger influence from environmental performance (0.154 points) and a higher adjusted $R^2$ of 15.8%, compared to the *ETR* (*Current_ETR* = 14.6% and *GAAP_ETR* = 6.8%).

**Table 10.** Alternative measures of tax avoidance; regression results for *EPILLAR* variable.

| **Dependent Variable:** | ***CURRENT_ETR*** | ***GAAP_ETR*** | ***TBTD*** |
|---|---|---|---|
| *EPILLAR* | −0.020 *** (−3.645) | −0.015 *** (−3.291) | 0.154 *** (4.958) |
| Controls | YES | YES | YES |
| Constant | −0.007 ** (−2.371) | −0.003 (−1.163) | 0.027 ** (2.705) |
| Observations | 1001 | 1004 | 1004 |
| Adjusted R-squared | 0.146 | 0.068 | 0.158 |
| Year Fixed Effect | YES | YES | YES |
| Industry Sector Fixed Effect | YES | YES | YES |
| Cluster by Industry Sectors | YES | YES | YES |

Table 10 presents the regression results after employing alternative measure of tax avoidance. *Current_ETR* and *GAAP_ETR* are the alternative proxies employed for the purpose of robustness check. The independent variable is environmental performance measured by *EPILLAR*. Robustness t-statistics are reported in parentheses next to the coefficients. Standard errors are corrected at the firm level. ***, and **, denote significance at the 1%, and 5% levels, respectively. Variable definitions are provided in the Appendix A.

4.4.6. An Alternative Measure of Environmental Performance

In [56], a distinctive methodology is employed to evaluate environmental performance, emphasising two pivotal variables pertinent to our research. The first identified variable is environmental innovation, commonly referred to as eco-innovation. This notion is rooted in the environmental pillar of innovation, resources, and emissions. For our analysis, we utilise the eco-innovation score obtained from Refinitiv Eikon (see Equation (9)), quantified on a scale ranging from 0 to 100. To represent this score in decimal format, we divide it by 100. This leads to

$$\begin{aligned} TBTD_{it} = {} & \beta_0^{(4)} + \gamma_1^{(4)} ECO_{INNOVATION\,it} + \beta_2^{(4)} ROA_{it} + \beta_3^{(4)} SIZE_{it} + \beta_4^{(4)} FCF_{it} + \beta_5^{(4)} LEV_{it} \\ & + \beta_6^{(4)} MTB_{it} + \beta_7^{(4)} AIN_{it} + \beta_8^{(4)} SG_{it} + \beta_9^{(4)} LIQ_{it} + \beta_{10}^{(4)} AGE_{it} + \beta_{11}^{(4)} BIG4_{it} \\ & + \beta_{12}^{(4)} BODSIZE_{it} + \beta_{13}^{(4)} BODIND_{it} + \beta_{14}^{(4)} FFIN_{it} + Industry\ Sector\ Fixed\ Effect \\ & + Year\ Fixed\ Effect + \varepsilon_{it}^{(4)} \end{aligned} \tag{9}$$

The second variable considered is environmental intensity (see Equation (10)). While Albitar et al. [56] quantify this variable by dividing environmental expenditure by total revenue, we adopt an alternative approach. We calculate research and development intensity as research and development expenditure divided by total assets, reflecting the organisation's dedication to improving environmental performance through research and development, i.e.,

$$\begin{gathered} TBTD_{it} = \beta_0^{(5)} + \gamma_2^{(5)} RND_{INTENS\,it} + \beta_2^{(5)} ROA_{it} + \beta_3^{(5)} SIZE_{it} + \beta_4^{(5)} FCF_{it} + \beta_5^{(5)} LEV_{it} + \\ \beta_6^{(5)} MTB_{it} + \beta_7^{(5)} AIN_{it} + \beta_8^{(5)} SG_{it} + \beta_9^{(5)} LIQ_{it} + \beta_{10}^{(5)} AGE_{it} + \beta_{11}^{(5)} BIG4_{it} + \beta_{12}^{(5)} BODSIZE_{it} + \\ \beta_{13}^{(5)} BODIND_{it} + \beta_{14}^{(5)} FFIN_{it} + Industry\ Sector\ Fixed\ Effect + Year\ Fixed\ Effect + \varepsilon_{it}^{(5)} \end{gathered} \tag{10}$$

The data in Table 11 indicate that both variables positively influence tax avoidance. Specifically, a higher degree of eco-innovation (see column 2 of Table 11, *ECO_INNOVATION* = 0.079) and a significantly increased investment in research and development (see column 3 of Table 11, *RND_INTENS* = 0.053) are linked to improved tax avoidance strategies among corporations. These findings are consistent with the overarching results concerning environmental performance.

**Table 11.** Alternative measures of environmental performance.

| | Dependent Variable *: TBTD* | | |
|---|---|---|---|
| | **(4)** | **(9)** | **(10)** |
| *EPILLAR* | 0.154 *** (4.958) | - | - |
| *ECO_INNOVATION* | - | 0.079 ** (2.487) | - |
| *RND_INTENS* | | | 0.053 * (1.909) |
| Controls | YES | YES | YES |
| Constant | 0.027 ** (2.705) | 0.020 ** (2.595) | 0.008 (0.691) |
| Observations | 1004 | 1004 | 1004 |
| Adjusted R-squared | 0.158 | 0.150 | 0.142 |
| Year Fixed Effect | YES | YES | YES |
| Industry Sector Fixed Effect | YES | YES | YES |
| Cluster by Industry | YES | YES | YES |

Table 11 presents the regression results after employing alternative measure of environmental performance. *ECO_INNOVATION* (environmental innovation score) and *RND_INTENS* (research and development intensity) are the alternative proxies employed for the purpose of robustness check. Robustness t-statistics are reported in parentheses next to the coefficients. Standard errors are corrected at the firm level. ***, **, and * denote significance at the 1%, 5%, and 10% levels, respectively.

## 5. DISCUSSION AND CONCLUSIONS

Our research examines the causal relationship between a firm's environmental performance and its tax-related strategies, with a particular focus on tax avoidance. By analysing a sample of FTSE All-Share companies from 2014 to 2022, our study demonstrates that enhanced environmental performance leads to a significant increase in tax avoidance, —our H1. This aligns with the earlier finding of [51], indicating that participation in environmental initiatives does not yield immediate financial benefits, leading managers to resort to tax avoidance to boost short-term earnings. The cross-sectional analysis indicates that this effect is particularly significant for firms experiencing higher financial constraints. Our study uses propensity score matching, entropy balancing, the Heckman model, and instrumental variable analysis to tackle the potential issues of correlated omitted variables, sample selection bias, and other endogeneity concerns. After conducting these analyses, the study results are consistent with the baseline findings. Moreover, the results remain robust even when alternative tax avoidance measures are applied.

Our study substantially contributes to the existing body of literature by offering several new insights into companies' environmental practices and tax compliance. Firstly, it broadens the investigation into the determinants of corporate tax avoidance behaviours [18,19,28] by utilising *TBTD* and *ETR* with the environmental aspect as an alternative to CSR and ESG. The findings indicate that the variables effectively predict tax avoidance, demonstrating that environmental performance impacts tax planning behaviour. Likewise, the authors of [28] demonstrate that enhanced CSR performance mitigates the extent of tax avoidance. Building upon these findings, this study reveals that firms extensively engaged in environmental initiatives are more inclined to adopt tax avoidance practices. This indicates that companies with greater environmental performance may employ tax avoidance strategies to alleviate the financial burdens associated with such initiatives.

Our study enriches the academic discourse by illuminating the connection between corporate financial health and tax behaviour, offering insights into how resource limitations shape compliance and strategic decision making within the context of environmental investments.

The findings of this research act as a benchmark, indicating that after the issuance of the 2008 Climate Change Act by the United Kingdom government, there has been a substantial increase in corporate involvement and performance, particularly concerning environmental initiatives. The increasing frequency of reported sample data each year evidence this augmentation. Nonetheless, this enhancement in environmental performance requires meticulous oversight from the UK government. Our findings suggest that improved environmental performance may be associated with a rise in tax avoidance strategies, an issue that warrants governmental scrutiny. To mitigate this concern, it would be prudent for the government to devise specific policies that reinforce environmental performance following environmental, social, and governance (ESG) criteria while concurrently remaining vigilant regarding tax avoidance practices. A recommendation for policymakers would be to design enforcement mechanisms on ESG compliance that account for firms' financial pressures. Tailoring these mechanisms to target industries or firm sizes particularly prone to tax evasion or aggressive tax planning can enhance compliance.

Additionally, introducing temporary tax relief or incentives for financially constrained firms could reduce the motivations for non-compliance, providing support while maintaining regulatory oversight (our H2). One possible suggestion to simultaneously boost investment in environmental initiatives and enhance eco-innovation while also maintaining resources needed for these initiatives is the widening of eco-tax policy, including the existing and new green taxes (HM Revenue & Customs (n.d.) *Climate Change Levy*. Available at: https://www.gov.uk/green-taxes-and-reliefs/climate-change-levy (accessed on: 24 December 2024)) (e.g., Climate Change Levy and Packaging Tax) and tailored green tax shifting policies. Within a global context, it has been shown that environmental taxes positively influence sustainability by incentivizing companies to curb carbon emissions, reduce overall pollution, and enhance environmental and economic efficiency [90,91], whereby country-specific economic contexts and enforcement policies can also serve as a key moderator, as in Bădîrcea et al. [92]. Sustained environmental investments positively influence long-term financial health (as opposed to short-term gains achieved, e.g., via tax avoidance) creating a virtuous cycle of sustainability and profitability as demonstrated in [89].

Therefore, it is vital to align the objectives of corporate managers and policymakers to ensure that pursuing ESG goals does not inadvertently give rise to additional challenges, such as tax avoidance, as the government strives towards its targets for 2050.

*Some Limitations and Suggestions for Future Research*

Lastly, we would like to highlight some limitations in our research and suggestions for future studies. First, we purposefully exclude financial and utility firms to refine our sample. Nonetheless, this exclusion limits a comprehensive understanding of the relationship between environmental performance and tax avoidance, particularly since these sectors exhibit unique financial structures, tax strategies, and regulatory constraints. Moreover, our study concentrates on the United Kingdom, whence indicating that the findings might not be universal or even generalizable to other countries with distinct tax policies, corporate governance frameworks, and sustainability priorities.

Thus, future research could explore the moderating roles of governance structures and international tax regimes, advancing our understanding of how institutional and regulatory contexts influence the relationship between environmental performance and tax avoidance. Specifically, examining how corporate tax strategies adapt in response to shifts in policy frameworks would be a promising area for future research. This would provide deeper insights into the dynamic interplay between regulatory environments and corporate behaviour, shedding light on how firms navigate evolving institutional pressures and adjust their strategies accordingly. For instance, research timelines might be delineated into two distinct intervals: before and after enforcing regulations. This would facilitate a thorough analysis of the effects of regulatory modifications, financial crises (such as the COVID-19 pandemic), and global sustainability initiatives.

Furthermore, we use a generalised proxy for environmental performance based on the environmental pillar, a widely accepted measure in academic research. On the other hand, this proxy may not fully capture all aspects of environmental performance, indicating the potential benefit of developing more nuanced measures in future research [93,94].

Additionally, comparisons may be drawn between corporations that exhibit environmental consciousness and those that do not. Cross-country analyses are also vital for examining tax avoidance behaviour in diverse contexts, encompassing developed and developing nations. One could point to findings highlighting tourism's potential as a sustainable industry with long-term growth prospects, like distinguishing countries with high levels of corruption but strong environmental awareness [95]. This insight might be of interest for policymakers aiming to balance economic development with environmental protection. This approach would help assess the global relevance of our findings and clarify regional variations in corporate tax avoidance practices.

In conclusion, our study contributes to the ongoing discussion on the challenges and strategies companies face in improving their ESG performance. It highlights the importance of implementing governmental policies that not only promote sustainable practices but also enhance transparency and accountability, particularly in financial reporting and tax planning.

**Authors Contributions:** Conceptualization, P.E.S., M.A. and P.K.; Methodology, P.E.S., M.A. and P.K.; Software, P.E.S.; Validation, P.E.S. and M.A.; Formal analysis, P.E.S., M.A. and P.K.; Investigation, P.E.S., M.A. and P.K.; Resources, P.E.S.; Data curation, P.E.S., M.A. and P.K.; Writing—original draft, P.E.S.; Writing—review & editing, P.E.S., M.A. and P.K. All authors have read and agreed to the published version of the manuscript.

**Funding:** This work by M.A. was partially supported by the project "A better understanding of socio-economic systems using quantitative methods from physics" funded by European Union---NextgenerationEU and Romanian Government, under National Recovery and Resilience Plan for Romania, contract no.760034/23.05.2023, code PNRR-C9-I8-CF 255/29.11.2022, through the Romanian Ministry of Research, Innovation and Digitalization, within Component 9, "Investment I8".

**Institutional Review Board Statement:** Not applicable

**Data Availability Statement:** The data presented in this study are available upon a valid request from the corresponding author.

**Conflicts of Interest:** The authors declare no conflict of interest.

## Appendix A

**Table A1.** Variable definitions.

| | Description | Source |
|---|---|---|
| **Dependent Variable:** *TBTD* | Total book-tax difference—a proxy to represent tax avoidance | Refinitiv Eikon |
| **Independent Variable:** | | |
| *EPILLAR* | Environmental performance—measured as the environmental performance pillar score | Refinitiv Eikon |
| **Control Variables:** | | |
| *ROA* | Return on assets—measured as income after taxes for the fiscal period divided by the average total assets | Refinitiv Eikon |
| *SIZE* | Firm size—measured by the natural logarithm of total assets | Refinitiv Eikon |
| *FCF* | Free cash flow—measured by cash flow from operations divided by total sales | Refinitiv Eikon |
| *LEV* | Leverage—measured by total debt divided by total assets | Refinitiv Eikon |
| *MTB* | Market-to-book ratio—measured by company market capitalisation divided by book value capitalisation | Refinitiv Eikon |
| *AIN* | Asset/income ratio—measured by total assets divided by net income before taxes | Refinitiv Eikon |
| *SG* | Sales growth—measured as the sales from year $t$ minus the sales from $t-1$ divided by the sales from $t-1$ | Refinitiv Eikon |
| *LIQ* | Liquidity—measured as current assets divided by current liabilities | Refinitiv Eikon |
| *AGE* | Firm age—measured by the natural logarithm of year t minus the date of incorporation plus 1 | Refinitiv Eikon |
| *BIG4* | Big 4—a dummy variable to denote whether the auditor is affiliated with one of the BIG 4 auditor firms (1) or not (0) | Refinitiv Eikon |
| *BODSIZE* | Board size—measured by the natural logarithm of the number of board members | Refinitiv Eikon |
| *BODIND* | Board independence—measured by the proportion of independent directors on the board | Refinitiv Eikon |
| *FFIN* | Financial Opacity—A measure of firm-level financial transparency is determined by industry and year adjusted total scaled accruals, based on previous research by Bhattacharya et al. [96]<br><br>The main variable for this model is a scaled accrual, which is an absolute value calculated using the formula $(\Delta CA + \Delta CL + \Delta CASH - \Delta STD + DEP + \Delta TP)/\text{lag}(TA)$, where $\Delta CA$ represents the change in total current assets, $\Delta CL$ represents the change in total current liabilities, $\Delta CASH$ represents the change in cash, $\Delta STD$ represents the change in the current portion of long-term debt included in total current liabilities, *DEP* represents depreciation and amortisation expense, $\Delta TP$ represents the change in income taxes payable, and lag(*TA*) represents total assets at the end of the previous year. *FFIN* takes the value of 1 if a firm has a higher than industry year mean of *ACCRUAL*, and 0 otherwise. | Refinitiv Eikon |

REFERENCES

1. House of Lords, 1st Report of Session 2013–14-Microsoft Word—Corporate Taxation Report Final Layout.doc. Available online: https://www.parliament.uk/globalassets/documents/lords-information-office/2015/hl-annual-report-2013-14.pdf (accessed on 24 December 2024).

2. Garner, B.A.; Black, H.C. Black's Law Dictionary, 9th ed.;West: St. Paul, MN, USA, 2009.

3. Souguir, Z.; Lassoued, N.; Khanchel, I.; Bouzgarrou, H. Environmental performance and corporate tax avoidance: Green-washing policy or eco-responsibility? The moderating role of ownership structure. J. Clean. Prod. 2024, 434, 140152.

4. Rego, S.O. Tax-avoidance activities of U.S. multinational corporations. Contemp. Account. Res. 2003, 20, 805–833.

5. Dyreng, S.D.; Hanlon, M.; Maydew, E.L. Long-run corporate tax avoidance. Account. Rev. 2008, 83, 61–82.

6. Ying, T.; Wright, B.; Huang, W. Ownership structure and tax aggressiveness of Chinese listed companies. Int. J. Account. Inf. Manag. 2017, 25, 313–332.

7. Crocker, K.J.; Slemrod, J. Corporate tax evasion with agency costs. J. Public Econ. 2005, 89, 1593–1610.

8. Armstrong, C.S.; Blouin, J.L.; Jagolinzer, A.D.; Larcker, D.F. Corporate governance incentives, and tax avoidance. J. Account. Econ. 2015, 60, 1–17.

9. Clarkson, P.M.; Li, Y.; Pinnuck, M.; Richardson, G.D. The valuation relevance of greenhouse gas emissions under the European Union carbon emissions trading scheme. Eur. Account. Rev. 2015, 24, 551–580.

10. Feng, C.; Zhu, X.; Gu, Y.; Liu, Y. Does the carbon emissions trading policy increase corporate tax avoidance? Evidence from China. Front. Energy Res. 2022, 9, 821219.

11. Khanchel, I.; Lassoued, N.; Bargaoui, I. Pollution Control Bonds and Environmental Performance in Energy Utility Firms: Is There an Incantation Effect? Int. J. Energy Sect. Manag. 2024, 18, 1066–1087.

12. Desai, M.A.; Dharmapala, D. Corporate tax avoidance and firm value. Rev. Econ. Stat. 2009, 91, 537–546.

13. Wang, F.; Xu, S.; Sun, J.; Cullinan, C.P. Corporate tax avoidance: A literature review and research agenda. J. Econ. Surv. 2020, 34, 793–811.

14. Aguilera, R.V.; Rupp, D.E.; Williams, C.A.; Ganapathi, J. Putting the S back in corporate social responsibility: A multilevel theory of social change in organizations. Acad. Manag. Rev. 2007, 32, 836–863.

15. Ntim, C.G.; Soobaroyen, T. Corporate governance and performance in socially responsible corporations: New empirical insights from a neo-institutional framework. Corp. Gov. Int. Rev. 2013, 21, 468–494.

16. Klassen, R.D.; Whybark, D.C. The impact of environmental technologies manufacturing performance. Acad. Manag. J. 1999, 42, 599–615.

17. Sarkis, J.; Cordeiro, J.J. An empirical evaluation of environmental efficiencies and firm performance: Pollution prevention versus end-of-pipe practice. Eur. J. Oper. Res. 2001, 135, 102–113.

18. Beck, T.; Lin, C.; Ma, Y. Why do firms evade taxes? The role of information sharing and financial sector outreach. J. Financ. 2014, 69, 763–817.

19. Dyreng, S.; Hanlon, M.; Maydew, E.L. The effects of executives on corporate tax avoidance. Account. Rev. 2010, 85, 1163–1189.

20. Na, K.; Yan, W. Languages and corporate tax avoidance. Rev. Account. Stud. 2022, 27, 148–184.

21. Lanis, R.; Richardson, G. Is corporate social responsibility performance associated with tax avoidance? J. Bus. Ethics 2015, 127, 439–457.

22. Ilinitch, A.Y.; Soderstrom, N.S.; Thomas, T.E. Measuring Corporate Environmental Performance. J. Account. Public Policy 1998, 17, 383–408.

23. Hassel, L.; Nilsson, H.; Nyquist, S. The value relevance of environmental performance. Eur. Account. Rev. 2005, 14, 41–61.

24. Law, K.K.; Mills, L.F. Taxes and financial constraints: Evidence from linguistic cues. J. Account. Res. 2015, 53, 777–819.

25. Haque, T.; Pham, T.P.; Yang, J. Geopolitical risk, financial constraints, and tax avoidance. J. Int. Financ. Mark. Inst. Money 2023, 88, 101858.

26. Alm, J.; Liu, Y.; Zhang, K. Financial constraints and firm tax evasion. Int. Tax Public Financ. 2019, 26, 71–102.

27. Mkadmi, J.E.; Ben Ali,W. How does tax avoidance affect corporate social responsibility and financial ratio in emerging economies? J. Econ. Criminol. 2024, 5, 100070.

28. Felicia, J.; Yusnaini. The Influence of ESG (Environmental, Social, and Governance) on the Performance of Tax Payments inTechnology Companies Listed on the IDX for the Period of 2017–2021. Daengku J. Humanit. Soc. Sci. Innov. 2023, 3, 640–647.

29. Lanis, R.; Richardson, G. Corporate social responsibility and tax aggressiveness: An empirical analysis. J. Account. Public Policy 2012, 31, 86–108.

30. Davies, A.K.; Guenther, D.A.; Krull, L.K.; Williams, B.M. Do Socially Responsible Firms Pay More Taxes? Account. Rev. 2016, 91, 47–68.

31. Gulzar, M.A.; Cherian, J.; Sial, M.S.; Bedulescu, A.; Thu, P.A.; Badulescu, D.; Khuong, N.V. Does corporate social responsibility influence corporate tax avoidance of Chinese listed companies? Sustainability 2018, 10, 4549.

32. Jiang, H.; Hu,W.; Jiang, P. Does ESG performance affect corporate tax avoidance? Evidence from China. Financ. Res. Lett. 2024, 61, 105056.

33. Laguir, I.; Staglianò, R.; Elbaz, J. Does corporate social responsibility affect corporate tax aggressiveness? J. Clean. Prod. 2015, 107, 662–675.

34. Abid, S.; Dammak, S. Corporate social responsibility and tax avoidance: The case of French companies. J. Financ. Rep. Account. 2022, 20, 618–638.

35. Edwards, A.; Schwab, C.; Shevlin, T. Financial Constraints and Cash Tax Savings. Account. Rev. 2016, 91, 859–881.

36. Penman, S. Financial Statement Analysis and Security Valuation, 5th ed.; McGraw-Hill Higher Education: New York, NY, USA, 2012.

37. Sun, J.; Makosa, L.; Yang, J.; Yin, F.; Sitsha, L. Does corporate tax planning mitigate financial constraints? Evidence from China. Int. J. Financ. Econ. 2021, 26, 4874–4894.

38. Leong, C.K.; Yang, Y.C. Constraints on "doing good": Financial constraints and corporate social responsibility. Financ. Res. Lett. 2021, 40, 101694.

39. Kabir, M.N.; Miah, M.D.; Rahman, S.; Alam, M.S.; Sarker, T. Financial constraints and Environmental Innovations: Evidence from Global Data. 2024.
Available online: https://papers.ssrn.com/sol3/papers.cfm?abstract_id=4848562 (accessed on
22 December 2024).

40. Ma, Z.; Liang, D.; Yu, K.H.; Lee, Y. Most cited business ethics publications: Mapping the intellectual structure of business ethics studies in 2001–2008. Bus. Ethics 2012, 21, 286–297.

41. Carroll, A.B. A three-dimensional conceptual model of corporate performance. Acad. Manag. Rev. 1979, 4, 497–505.

42. Baden, D. A reconstruction of Carroll's pyramid of corporate social responsibility for the 21st century. Int. J. Corp. Soc. Responsib. 2016, 1, 1–15.

43. Carroll, A.B. The pyramid of corporate social responsibility: Toward the moral management of organisational stakeholders. Bus. Horiz. 1991, 34, 39–48. Available online:
https://link.gale.com/apps/doc/A11000639/AONE?u=anon~cf135a41&sid=googleScholar&xid=4b0a43c4 (accessed on 24 December 2024).

44. Rehman, I.U.; Shahzad, F.; Hanif, M.A.; Arshad, A.; Sergi, B.S. Financial constraints and carbon emissions: An empirical investigation. Soc. Responsib. J. 2024, 20, 761–782.

45. Huseynov, F.; Klamm, B.K. Tax avoidance, tax management and corporate social responsibility. J. Corp. Financ. 2012, 18, 804–827.

46. Jones, S.; Baker, M.; Ben, F.L. The relationship between CSR and tax avoidance: An international perspective. Aust. Tax Forum 2017, 32, 95–127.

47. Kiesewetter, D.; Manthey, J. Tax avoidance, value creation and CSR—A European perspective. Corp. Gov. 2017, 17, 803–821.

48. Alsaadi, A. Financial-tax reporting conformity, tax avoidance and corporate social responsibility. J. Financ. Report. Account. 2020, 18, 639–659.

49. Muller, A.; Kolk, A. Responsible Tax as Corporate Social Responsibility: The Case of Multinational Enterprises and Effective Tax in India. Bus. Soc. 2015, 54, 435–463.

50. Zeng, T. Relationship between corporate social responsibility and tax avoidance: International evidence. Soc. Responsib. J. 2019, 15, 244–257.

51. Salhi, B.; Riguen, R.; Kachouri, M.; Jarboui, A. The mediating role of corporate social responsibility on the relationship between governance and tax avoidance: UK common law versus French civil law. Soc. Responsib. J. 2020, 16, 1149–1168.

52. Ortas, E.; Gallego-Álvarez, I. Bridging the gap between corporate social responsibility performance and tax aggressiveness: The moderating role of national culture. Account. Audit. Account. 2020, 33, 825–855.

53. Feng, Z.;Wang, Y.;Wang,W. Corporate carbon reduction and tax avoidance: International evidence. J. Contemp. Account. Econ. 2024, 20, 100416.

54. Faulkender, M.;Wang, R. Corporate Financial Policy and the Value of Cash. J. Financ. 2006, 61, 1957–1990.

55. Almeida, H.; Campello, M.; Weisbach, M.S. The Cash Flow Sensitivity of Cash. J. Financ. 2004, 59, 1777–1804.

56. Ferdous, L.T.; Rana, T.; Yeboah, R. Decoding the impact of firm-level ESG performance on financial disclosure quality. Bus. Strategy Environ. 2024, 34, 162–186.

57. Albitar, K.; Al-Shaer, H.; Liu, Y.S. Corporate commitment to climate change: The effect of eco-innovation and climate governance. Res. Policy 2023, 52, 104697.

58. Bayar, O.; Huseynov, F.; Sardarli, S. Corporate Governance, Tax Avoidance, and Financial Constraints. Financ. Manag. 2018, 47, 651–677.

59. Whited, T.M.;Wu, G. Financial Constraints Risk. Rev. Financ. Stud. 2006, 19, 531–559.

60. Alharbi, S.S.; Atawnah, N.; Ali, M.J.; Eshraghi, A. Gambling culture and earnings management: A novel perspective. Int. Rev. Econ. Financ. 2023, 86, 520–539.

61. Li, B.; Liu, Z.;Wang, R. When dedicated investors are distracted: The effect of institutional monitoring on corporate tax avoidance. J. Account. Public Policy 2021, 40, 106873.

62. Chung, S.G.; Goh, B.W.; Lee, J.; Shevlin, T. Corporate Tax Aggressiveness and Insider Trading. Contemp. Account. Res. 2019, 36, 230–258.

63. Amin, M.R.; Akindayomi, A.; Sarker, M.S.R.; Bhuyan, R. Climate policy uncertainty and corporate tax avoidance. Financ. Res. Lett. 2023, 58, 104581.

64. Delgado, F.J.; Fernández-Rodríguez, E.; García-Fernández, R.; Landajo, M.; Martínez-Arias, A. Tax avoidance and earnings management: A neural network approach for the largest European economies. Financ. Innov. 2023, 9, 19–25.

65. D'Amico, E.; Coluccia, D.; Fontana, S.; Solimene, S. Factors Influencing Corporate Environmental Disclosure. Bus. Strategy Environ. 2016, 25, 178–192.

66. Lucut Capras, I.; Achim, M.V.; Mara, E.R. Is tax avoidance one of the purposes of financial data manipulation? The case of Romania. J. Risk Financ. 2024, 25, 588–601.

67. Abdou, H.A.; Ellelly, N.N.; Elamer, A.A.; Hussainey, K.; Yazdifar, H. Corporate governance and earnings management nexus: Evidence from the UK and Egypt using neural networks. Int. J. Financ. Econ. 2021, 26, 6281–6311.

68. Thanasas, G.; Filiou, A.; Smaraidos, V. The impact of corporate governance on earnings management in emerging economies: The Greek evidence. Int. J. Comp. Manag. 2018, 1, 317-330.

69. Velte, P. Sustainable institutional investors, corporate sustainability performance, and corporate tax avoidance: Empirical evidence for the European capital market. Corp. Soc. Responsib. Environ. Manag. 2023, 30, 2406–2418.

70. Dhaliwal, D.S.; Radhakrishnan, S.; Tsang, A.; Yang, Y.G. Nonfinancial Disclosure and Analyst Forecast Accuracy: International Evidence on Corporate Social Responsibility Disclosure. Account. Rev. 2012, 87, 723–759.

71. Bhattacharya, U.; Daouk, H.; Welker, M. The World Price of Earnings Opacity. Account. Rev. 2003, 78, 641–678.

72. DeFond, M.L.; Hung, M. An empirical analysis of analysts' cash flow forecasts. J. Account. Econ. 2003, 35, 73–100.

73. Leuz, C.; Nanda, D.; Wysocki, P.D. Earnings management and investor protection: An international comparison. J. Financ. Econ. 2003, 69, 505–527.

74. Elamer, A.A.; Boulhaga, M.; Ibrahim, B.A. Corporate tax avoidance and firm value: The moderating role of environmental, social, and governance (ESG) ratings. Bus. Strategy Environ. 2024, 33, 7446–7461.

75. Hanlon, M.; Heitzman, S. A review of tax research. J. Account. Econ. 2010, 50, 127–178.

76. Schauer, C.; Elsas, R.; Breitkopf, N. A new measure of financial constraints applicable to private and public firms. J. Bank. Financ. 2019, 101, 270–295.

77. Lee, C.; Wang, C. Firms' cash reserve, financial constraint, and geopolitical risk. Pac. Basin Financ. J. 2021, 65, 101480. 

78. Baker, M.; Stein, J.C.; Wurgler, J. When Does the Market Matter? Stock Prices and the Investment of Equity-Dependent Firms. Q. J. Econ. 2003, 118, 969–1005.

79. Hasan, I.; Hoi, C.K.S.; Wu, Q.; Zhang, H. Beauty is in the eye of the beholder: The effect of corporate tax avoidance on the cost of bank loans. J. Financ. Econ. 2014, 113, 109–130.

80. Panousi, V.; Papanikolaou, D. Investment, Idiosyncratic Risk, and Ownership. J. Financ. 2012, 67, 1113–1148.

81. Li, D. Financial Constraints, R&D Investment, and Stock Returns. Rev. Financ. Stud. 2011, 24, 2974–3007.

82. Duchin, R. Cash Holdings and Corporate Diversification. J. Financ. 2010, 65, 955–992.

83. Kaplan, S.N.; Zingales, L. Do Investment-Cash Flow Sensitivities Provide Useful Measures of Financing Constraints? Q. J. Econ. 1997, 112, 169–215.

84. Lamont, O.; Polk, C.; Saá-Requejo, J. Financial Constraints and Stock Returns. Rev. Financ. Stud. 2001, 14, 529–554.

85. Hainmueller, J. Entropy Balancing for Causal Effects: A Multivariate Reweighting Method to Produce Balanced Samples in Observational Studies. Political Anal. 2012, 20, 25–46.

86. Shannon, C.E. A mathematical theory of communication. Bell Syst. Tech. J. 1948, 27, 623–656.

87. Chen, C.; Huang, T.; Garg, M.; Khedmati, M. Governments as customers: Exploring the effects of government customers on supplier firms' information quality. J. Bus. Financ. Account. 2021, 48, 1630–1667.

88. Khan, M.; Srinivasan, S.; Tan, L. Institutional Ownership and Corporate Tax Avoidance: New Evidence. Account. Rev. 2017, 92, 101–122.

89. Chyz, J.A.; Ching Leung, W.S.; Zhen Li, O.; Meng Rui, O. Labor unions and tax aggressiveness. J. Financ. Econ. 2013, 108, 675–698.

90. HM Revenue & Customs. Climate Change Levy. Available online: https://www.gov.uk/green-taxes-and-reliefs/climate-changelevyn (accessed on 24 December 2024).

91. Bigerna, S.; D'Errico, M.C.; Micheli, S.; Polinori, P. Environmental-economic efficiency for carbon neutrality: The role of eco-innovation, taxation, and globalization in OECD countries. Appl. Econ. 2024, 56, 3568–3581.

92. Gao, X.; Fan, M. Environmental taxes, eco-innovation, and environmental sustainability in EU member countries. Environ. Sci. Pollut. Res. 2023, 30, 101637–101652.

93. Bădîrcea, R.M.; Florea, N.M.; Manta, A.G.; Puiu, S.; Doran, M.D. Comparison between Romania and Sweden Based on Three Dimensions: Environmental Performance, Green Taxation and Economic Growth. Sustainability 2020, 12, 3817.

94. Bassetti, T.; Blasi, S.; Sedita, S.R. The management of sustainable development: A longitudinal analysis of the effects of environmental performance on economic performance. Bus. Strategy Environ. 2021, 30, 21–37.

95. Arsawan, I.W.E.; Hariyanti, N.K.D.; Azizah, A.; Suryantini, N.P.S.; Darmayanti, N.P.A. Internet of things towards environmental performance: A scientometrics and future research avenues. E3S Web Conf. 2024, 501, 01011.

96. Wang, M.C.; Chen, Z. The relationship among environmental performance, R&D expenditure, and corporate performance: Using simultaneous equations model. Qual. Quant. 2022, 56, 2675–2689.

97. Cozma, A.-C.; Achim, M.C.; Mare, C.; Coros, M.M. The Moderating Role of Tourism in the Impact of Financial Crime on Deforestation. J. Clean. Prod. 2025, 486, 144475.